\documentclass[%
nofootinbib,
10pt,
 amsmath,amssymb,
 longbibliography,
 aps,
 prd, 
]{revtex4-2}

\usepackage{graphicx}
\usepackage{dcolumn}
\usepackage{bm}
\usepackage{caption}
\usepackage{subcaption} 

\usepackage{amsmath}
\usepackage[toc,page]{appendix}
\usepackage{lipsum}
\usepackage{threeparttable}
\usepackage[bottom]{footmisc}

\usepackage{hyperref}

\usepackage{cleveref}

\begin{document}

\preprint{APS/123-QED}

\title{The optimal redshift for dark energy II: application to cosmological data and the evidence for the phantom crossing of the CPL equation of state}

\author{Travis Seth Rippentrop}
\email{tsr200000@utdallas.edu}
\author{Mustapha Ishak}
\email{mishak@utdallas.edu}
\author{Kristian Gonzalez}
\email{keg180005@utdallas.edu}

\affiliation{Department of Physics, The University of Texas at Dallas, Richardson, TX 75080, USA}

\date{\today}

\begin{abstract}
     Recent results from DESI and other cosmological datasets have indicated a preference for dynamical dark energy, with a time-evolving equation of state (EOS), denoted by $w(z)$. Furthermore, analyses using the commonly used CPL parameterization give an EOS that undergoes a crossing of the $w(z)=-1$ phantom line. In a companion paper-I, we assumed the CPL parameterization and its generalization, and introduced the formalism of the optimal scale factor (or redshift), a quantity designed to simultaneously maximize the distance from the cosmological constant value of $-1$ and to minimize the uncertainty on the EOS, thereby maximizing the tension estimation with the cosmological constant at that specific redshift for a given dataset combination. In this paper-II, we apply the optimal-redshift formalism to currently available baryon acoustic oscillation (BAO), Cosmic Microwave Background (CMB), and Type-Ia supernova (SN) dataset combinations, with a particular focus on the phantom-line crossing predicted within the CPL parameterization. Motivated by the relatively low significance of previous constraints on the EOS in the region below the phantom line, we calculate the optimal redshift for a variety of dataset combinations in order to identify those that best constrain departures from $w(z)=-1$ before and after the crossing. We find dataset combinations and corresponding optimal redshifts for which $w(z)$ lies below the ``$-1$'' line with new significance levels reaching $3.01$--$3.22\sigma$ before the crossing point, and $3.30$--$3.55\sigma$ above ``$-1$" after the crossing. Our focus in this work is the application of a new framework that maximizes the statistical significance of the phantom-crossing signal within CPL parameterizations using currently available datasets. Determining whether this crossing is effective or intrinsic, and identifying the underlying microphysical models, constitute separate questions from the aim of the present work and remain important directions for future investigation, further motivated by our findings. 

\end{abstract}

\maketitle

\section{Introduction \label{section:intro}}

The standard $\Lambda$CDM model of cosmology has enjoyed a long period of success when compared to observations over the past few decades. The model assumes a universe consisting of $\sim 5\%$ baryonic matter, $\sim25\%$ dark matter, and $\sim 70 \%$ constant dark energy in the form of a cosmological constant. However, recent baryon acoustic oscillation (BAO) measurements from the Dark Energy Spectroscopic Instrument (DESI), when combined with cosmic microwave background (CMB) data and supernova surveys, have challenged this prevailing view by preferring a dynamical dark energy (DDE) with a time-evolving equation of state (EOS), denoted as $w(z)$, where $z$ is the redshift see, e.g. \cite{DESI:2024mwx, DESI:2025zpo, ishak2024desi, DESI:2025zgx,DESI:2025fii,DESI:2025gwf}. This finding was confirmed by a number of follow up studies and has led to a significant amounts of recent literature regarding a plethora of alternative dark energy models as well as examinations of observational constraints and possible systematic effects. An incomplete list of these works is given in \cite{Tada:2024xau, Wang:2024qan, Yin:2024xau, Luongo:2024xau, Cortes:2024xau, Colgain:2024xqj, Wang:2024rjd, Berghaus:2024kra, Wang:2024xau, Wang:2024pui, Shlivko:2024llw, Dinda:2024kjf, Bhattacharya:2024xau, Ramadan:2024xau, Roy:2024kni, Gialamas:2024lyw, Notari:2024xau, Liu:2024gfy, Orchard:2024xau, Hernandez-Almada:2024xau, Pourojaghi:2024xau, Giare:2024gpk, Jiang:2024xau, Efstathiou:2024xau, Reboucas:2024xau, RoyChoudhury:2024xau, Dhawan:2024xau, Linder:2024rdj, Park:2024xau2, Notari:2024xau2, Gao:2024ily, Fikri:2024klc, Tiwari:2024gzo, Tang:2024lmo, Zheng:2024xau, Odintsov:2024xau, Colgain:2024mtg, Lewis:2024xau, Koussour:2024xau, Sakr:2025fay, Yang:2025xau, Huang:2025xau, Wolf:2025xau, Giare:2025pzu, Sousa-Neto:2025gpj, Ormondroyd:2025exu, Khoury:2025txd, Ormondroyd:2025iaf, Brandenberger:2025hof, Nesseris:2025lke, Kessler:2025kju, You:2025uon, Shlivko:2025fgv, Cai:2025mas, Popovic:2025glk, Li:2025ops, Kou:2025yfr, Santos:2025wiv, DESI:2025wyn, Akarsu:2025dmj, Dinda:2025iaq, Wang:2025bkk, Mirpoorian:2025rfp, RoyChoudhury:2025dhe, Scherer:2025esj, Liu:2025mub, Chen:2025mlf, Chen:2025wwn, Efstathiou:2025tie, vanderWesthuizen:2025iam, Sabogal:2025jbo, Linder:2025zxb, Odintsov:2025jfq, Araya:2025rqz, Li:2025eqh, Herold:2025hkb, Lee:2025kbn, Chen:2025jnr, Hogas:2025ahb, Qiang:2025cxp, Lee:2025pzo, Chudaykin:2025aux, Silva:2025twg, Camarena:2025upt, Chaudhary:2025pcc, Wang:2025vtw, Yao:2025wlx, Arora:2025msq, Paul:2025wix, Toomey:2025xyo, Fazzari:2025lzd, Dinda:2025hiu, Park:2025fbl, RoyChoudhury:2025iis, Alam:2025epg, Artola:2025zzb, Reeves:2025xau, Yadav:2025vgo, Rezaei:2025vhb, Xu:2025nsn, Bolotin:2013jpa, Wang:2016lxa, Wang:2024vmw, Spec-S5:2025uom, Figueruelo:2026eis, Li:2025muv, Li:2024qso, Guedezounme:2025wav, Antusch:2026ldp, Bedroya:2025fwh, Andriot:2026lac, Anchordoqui:2026hys, Sailer:2025lxj, Jhaveri:2025neg, Elbers:2025xvk, Hergt:2026moc}. Additionally, see \cite{li2026dark} for a recent review.  

Intriguingly, analyses based on phenomenological parameterizations of the EOS appear to indicate a phantom crossing, in which the dark energy EOS was in the phantom regime in the past (i.e., $w < -1$) and later transitioned to the region where $w > -1$ \cite{shlivko2026phantom}. Such a crossing also corresponds to a feature (or bump) in the dark energy density evolution where the density is maximized in the past and decreases in late times. This behavior is entirely unexpected within standard explanations of dynamical dark energy, such as single field models like quintessence \cite{vikman2005can}, and may require more complex models involving, for example, multi-field scalar components see, e.g. \cite{feng2005dark, Cai:2025mas}, an interacting dark sector see, e.g. \cite{Guedezounme:2025wav}, or modifications to gravity see, e.g. \cite{cai2026quintom, Nojiri:2025low} to explain it.

Previous studies using functional or binned parameterizations of the EOS have came short of providing significance levels exceeding $3\sigma$ for measurements below the phantom boundary of $w = -1$  \cite{DESI:2025fii}.

In this work, regardless of whether this crossing is intrinsic or effective \cite{li2026dark}, we provide a new framework that is able to extract a significance level of phantom behavior above the $3\sigma$ level before the crossing using only currently available datasets. As discussed in our companion paper-I \cite{paper-I}, we introduce the notion of the \emph{optimal redshift} (\emph{optimum} for short), a redshift that best minimizes the error on $w(z)$ while simultaneously maximizing the distance between $w(z)$ and the $w(a)=-1$ line, thereby maximizing the tension estimation with the cosmological constant value for a given dataset. This is defined within the CPL parameterization \cite{Chevallier:2001,Linder2003} and some generalization of it around any arbitrary redshift.  
Using this optimum we give evidence for deviation away from $w(a)=-1$ in both directions using different datasets combinations involving BAO, CMB, and Supernovae, including recently re-calibrated supernova datasets. From this deviation we show that there is statistically significant evidence for an intrinsic or effective crossing between the regime where $w(a)<-1$ to $w(a)>-1$ in the cosmic evolution given CPL parametrization. 

This paper-II is organized as follows. In \cref{section:methodology}, we lay out the methodology for the concept of a phantom crossing, review the notion of the optimal redshift, and discuss different methods of estimating tension with $\rm \Lambda CDM$ and the $w=-1$ line. In \cref{section:data}, we list the datasets used in this paper. In \cref{section:results}, we apply the formalism of the optimum to current datasets and run Monte-Carlo-Markov-Chains (MCMC) results for 15 different dataset combinations. And finally, we conclude with \cref{section:conclusion} while analyzing the implications of the evidence provided for a crossing using the optimal redshift within the CPL parameterization.

\section{Methodology \label{section:methodology}}
\subsection{The Phantom Crossing}
The equation of state of a substance is the relation between its energy density and its pressure and can be expressed in terms of $w$ where\footnote{We use $c=1$ throughout.}
\begin{equation}
    P=w \rho.
\end{equation}
The density of a substance with a certain value of $w$ will evolve as scale factor $a$ evolves according to the continuity equation,
\begin{equation}
    \dot{\rho} + 3H(1+w)\rho = 0.
\end{equation}

The $\rm \Lambda CDM$ model considers a dark energy density which is constant ($\dot{\rho}_{\rm DE}=0$) implying $w=-1$. However, this is only the simplest form of dark energy and substances with a non constant $w(a)$ are routinely considered. 

The most common parameterization of a variable equation of state is the Chevallier-Polarski-Linder (CPL) parameterization which is currently used in almost every report of cosmological experiments. It can be expressed as
\begin{equation}
    w(a) = w_0 + w_a(1-a).
\end{equation}
where scale factor $a$ is related to the redshift by $a=\frac{1}{1+z}$.

The $\rm \Lambda CDM$ model can be expressed in CPL by setting $w_0=-1$ and $w_a=0$. However, any deviation in $w_0$ or $w_a$ from these values indicates dynamical dark energy.

Recent cosmological parameter estimations have significantly hinted towards not only dynamical dark energy, but also a phantom crossing. A phantom crossing occurs when a variable equation of state changes from $w(a) > -1$ to $w(a)<-1$ or vise versa. 

The equation of state of a quintessence field with potential $V(\phi)$ is given by
\begin{equation}
    w = \frac{P}{\rho} = \frac{\frac12 \dot\phi-V(\phi)}{\frac 12 \dot\phi+V(\phi)}.
\end{equation}
Such a field can only produce $w\geq-1$. To make $w<-1$, a phantom field with a negative kinetic energy must be used instead of a quintessence field. However, if only a phantom field is used $w>-1$ is now forbidden. A phantom crossing in the universe's past is known as the quintom scenario and would require either both field types to effectively model or a more complicated form of a single field model. A quintom scenario could imply the need for much more complex physics than a simpler quintessence model to explain dynamical dark energy. Thus, thoroughly exploring the evidence for a phantom crossing is crucial to guide future exploration of cosmological models.

The scale factor at the time of the crossing can be determined using the CPL parameters,
\begin{equation}
    a_c=1+\frac{w_0+1}{w_a}.
    \label{ac definition}
\end{equation}
Current cosmological parameter estimation for CPL consistently estimates $w_0<-1$ which implies the condition for a quintom scenario to be $w_a < -(1+w_0)$. Typically, the crossing occurs around $a_c\sim0.7$ ($z_c\sim0.4$). We propose investigating the evidence for this crossing by noting the value of the CPL modeled $w(a)$ at two different scale factors/redshifts, $w_1=w(a_1)$ and $w_2=w(a_2)$ such that $w_1 > -1$ and $w_2<-1$. Since $w(a)$ is continuous in CPL, from the Intermediate Value Theorem (IVT), it is therefore clear that if $a_1$ and $a_2$ can be adequately shown to exist, there must be an $a_c$ between them such that $w(a_c)=-1$, constituting a crossing of the $w=-1$ line. However, uncertainties in the data can make the true existence of this crossing hard to demonstrate as the error bars on $w_1$ and $w_2$ are prone to overlap to a significant degree.

\subsection{The Optimum \label{subsection: optimum}}

We briefly review the optimal redshift formalism as presented in \cite{paper-I}. We start here though by considering a parameterization of $w(a)$ to be a first order Taylor expansion centered at $a_*$
\begin{equation}
    w(a)=w_*+w'(a-a_*).
\end{equation}
We may express these new parameters $w_*$ and $w'$ in terms of the usual CPL parameters as
\begin{equation}
    w_*=w_0+w_a+w'a \quad {\rm and} \quad w'=-w_a.
\end{equation}
Since $w'$ is simply a sign flipped version of $w_a$, we may re-express the Taylor expansion as
\begin{equation}
    w(a)=w_*+w_a(a_*-a)
\end{equation}
reusing the same $w_a$ parameter as CPL for mathematical and computational convenience. As in \cite{paper-I}, we refer to this parameterization as CPL$_*$, a generalization CPL and of which CPL is the special case when $a_*=1$. Parameter $w_*$ may now be expressed as
\begin{equation}
    w_*=w_0+w_a(1-a_*).
    \label{w* def}
\end{equation}
Note that $w_*$ is equivalent to $w(a)$ evaluated at $a_*$.

When performing cosmological parameter estimation using this parameterization, we may choose $a_*$ to fulfill our goal of finding a suitable $w_1$ and $w_2$. Interestingly, when changing $a_*$, the variance in $w_*$, as well as the covariance between $w_*$ and $w_a$, will change. A well explored example of this is the pivot. When $w_0$ and $w_a$ are estimated in CPL, they are typically anti-correlated. This causes their one dimensional uncertainties to be larger than if they were uncorrelated. The $a_*$ value that minimizes the variance in the value of $w_*$ is considered to be the pivot point, $a_p$ and $w(a_p)$ is denoted $w_p$.

Throughout this paper, angled brackets $\langle \rangle$ will denote an expectation value, thus the variance of quantity X can be expressed as
\begin{equation}
    {\rm var}(X)=\langle X^2\rangle-\langle X\rangle^2=\langle\delta X^2\rangle.
\end{equation}
Minimizing the variance of $w_*$, $\langle\delta w_*^2\rangle = \langle(\delta w_0 + (1-a_*) \delta w_a)^2\rangle$ gives \cite{Albrecht:2006um}
\begin{equation}
a_p = 1+\frac{\langle\delta w_a \delta w_0 \rangle}{\langle\delta w_a^2\rangle}.
    \label{ap definition}
\end{equation}
This value of $a_p$ also de-correlates $w_p$ and $w_a$ entirely by making the covariance $\langle \delta w_p \delta w_a\rangle =0$.

\begin{figure}
    \centering
    \includegraphics[width=0.7\linewidth]{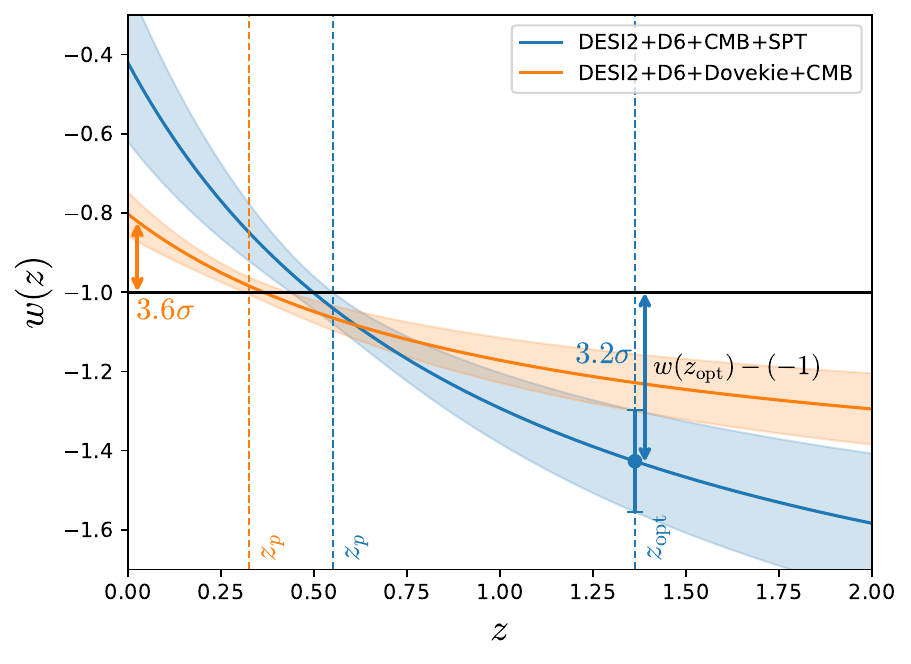}
    \caption{Plot of equation of state $w(z)$. The two dataset combinations with the most significant optimum tension are plotted. Shaded regions denote 1-sigma error. Dashed vertical lines denote either the pivot or the optimum of the dataset. The DESI2 + D6 + CMB + SPT dataset provides an optimum below $w=-1$ (before the crossing) with a $3.2\sigma$ tension. The DESI2 + D6 + Dovekie + CMB dataset has an optimum in the future which is not plotted. Instead, the $3.6\sigma$ tension of the present day value is used (after the crossing). Note that $z_p$ occurs too close to the crossing where even the $1\sigma$ error bars overlap $w=-1$, which is not the case at $z_{\rm opt}$.}
\end{figure}

Taking $a_1$ and $a_2$ near $a_p$ appears to be a good choice to demonstrate the existence of a crossing, as we minimize the error bars as much as possible. Unfortunately, $a_p$ for typical datasets is usually quite close to $a_c$. This proves undesirable for our purposes, as when $a_*$ is near $a_c$, $w_*$ is very nearly $-1$, so despite having minimized the error bars, we are also minimizing the difference between $w(a_*)=w_*$ and the $-1$ equation of state of a cosmological constant. It becomes extremely difficult at the pivot point to effectively determine if two $w$ values are on different sides of the crossing.

As presented in \cite{paper-I}, instead of minimizing variance, we may instead choose to minimize the error bars on $w_*$ while still best maximizing the distance between $w_*$ and $-1$. In other words, maximizing the ratio between the distance $w_*+1$ and the error bar on $w_*$,  which is analytically equivalent to minimizing the 1-dimensional $\Delta\chi^2$ between $w_*$ and $-1$,
\begin{equation}
   \frac{d}{da_*}\Delta\chi^2_{\rm1D}=-\frac{d}{da_*}\frac{(w_*+1)^2}{\langle\delta w_*^2\rangle} =0.
\end{equation}
This will yield the point of greatest possible tension between $w_*$ and $-1$, given the data. We call the $a_*$ value of this point the \emph{optimum}. It is given by
\begin{equation}
    a_{\rm opt} = \frac{w_a\big(\langle \delta w_0^2\rangle+\langle \delta w_0\delta w_a\rangle\big)-(w_0+1)\big(\langle \delta w_a^2\rangle+\langle \delta w_0\delta w_a\rangle\big)}{w_a\langle \delta w_0\delta w_a\rangle-(w_0+1)\langle \delta w_a^2\rangle}.
    \label{optimum definition}
\end{equation}

As shown in \cite{paper-I}, by using the definitions of $a_c$ \cref{ac definition} and $a_p$ \cref{ap definition} this equation can be re-written as
\begin{equation}
    a_{\rm opt} = a_p+\frac{(a_p-1)^2-\frac{\langle \delta w_0^2\rangle}{\langle \delta w_a^2\rangle}}{a_c-a_p}
\end{equation}
The $\Delta\chi^2_{\rm1D}$ value at the optimum is given by
\begin{equation}
    \Delta\chi_{\rm opt}^2 = -\frac{w_a^2}{\langle \delta w_a^2\rangle}\Bigg(1+\frac{(a_c-a_p)^2}{\frac{\langle \delta w_0^2\rangle}{\langle \delta w_a^2\rangle}-(a_p-1)^2}\Bigg).
    \label{chi opt}
\end{equation}
and the square root of its absolute value represents the maximum tension $\sigma_{\rm opt}$ between $w_*$ and $-1$ that can be achieved with a choice of $a_*$.

Each dataset and combination of datasets will produce its own $w_0$, $w_a$ estimates and covariance matrix meaning there is a unique $a_p$, $a_c$, and $a_{\rm opt}$ for each one. Datasets combinations which yield a $w_{\rm opt} > -1$ will serve as $w_1$ and will be referred to as a "type 1 combination". Likewise, "type 2 combinations" will have $w_{\rm opt}< -1$ and will serve as $w_2$. Analyzing the difference between the $w$ values at these optima as well as the difference between each $w_{\rm opt}$ and -1 will be crucial for our goals.

It is worth noting that, in general, $w_1$ and $w_2$ belong to two different functions (the best fit CPL $w(a)$ model for each dataset combination), and thus the IVT, as mentioned in section A, cannot be applied directly. However, since CPL has two parameters, there must exist a third CPL function, $w_{\rm IVT}(a)$, with it's own $w_0$ and $w_a$ which intersects both of these values. It is this function which we may apply the IVT to and determine the existence of an $a_c$ such that $w_{\rm IVT}(a_c)=-1$.

Including more high redshift data in a dataset gives a lower $a_p$ and vise versa. So, in general, datasets which prominently feature the CMB typically have optimums before the crossing and datasets that are heavy in Supernovae have optimums after the crossing.

One feature of this method to be cautious about is the fact that, since CPL is a linear parameterization, it could predict $w(a)$ values for hypothetical times either in the future with ``$a>1$'' or ``before" the Big Bang with nonphysical values. Likewise, the optimum scale factor values can be calculated at any scale factor, even ones in the future. We find that this commonly arises when combining BAO, Supernovae, and CMB where optimum scale factors are greater than 1. However, it is worth noting that even in this case, the maximal tension associated with the optimum is only slightly greater than the tension associated with today's CPL parameter $w_0$. Thus, instead of using the future optimum value, we choose to use the present $w_0$ as our $w_1$.

Additionally, in the limiting case $a_p \rightarrow a_c$, $a_{\rm opt}$ will be far in the future or well ``before" the Big Bang and $\Delta \chi^2_{\rm opt}$ is minimized ($\Delta \chi^2_{\rm opt}\rightarrow\frac{w_a^2}{\langle\delta w_a^2\rangle}$).

\begin{figure*}    \centering
    \includegraphics[width=1\textwidth]{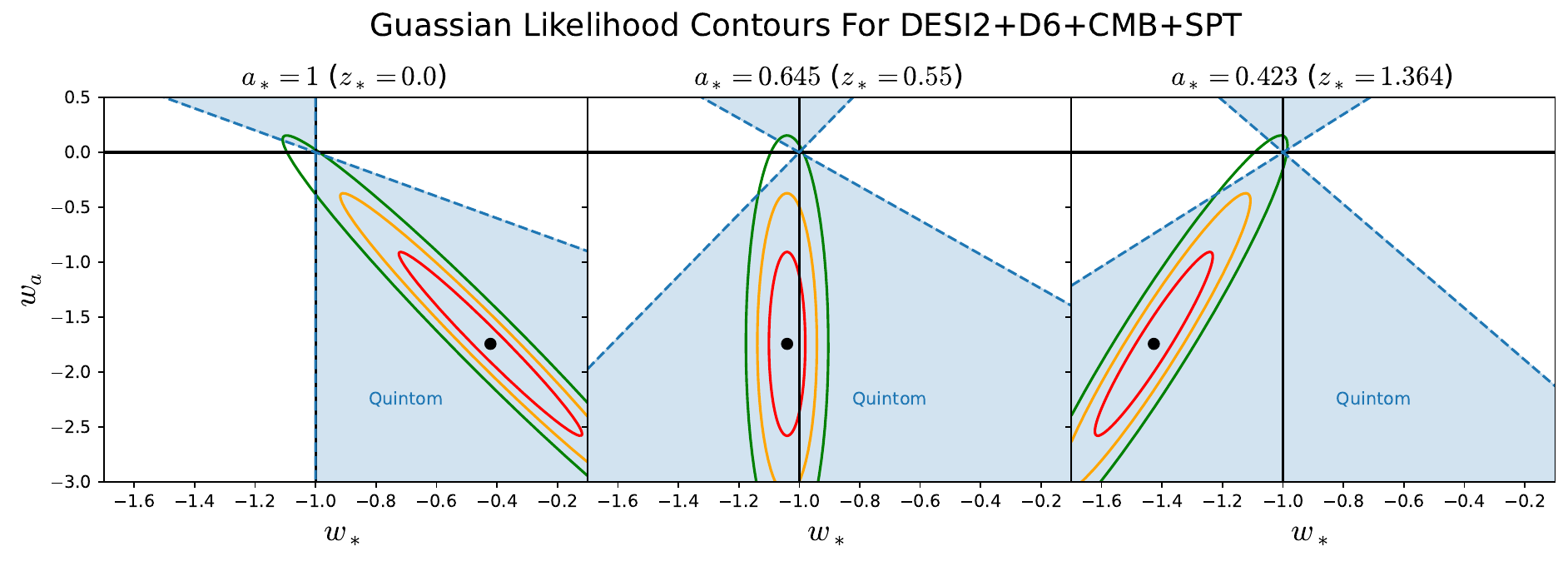}
    \hfill
     \caption{Estimated n-sigma Contours at Different $a_*$. Using CPL parameters for the DESI2 + D6 + CMB + SPT dataset combination and \cref{cpl approx chi2} predictions of likelihood contours are made for the present, the pivot, and the optimum respectively. Red, orange, and green correspond to 1, 2, and 3 sigma contours respectively, while the black point denotes the mean. In general, as $a_*$ decreases ($z_*$ increases), the ellipse rotates clockwise. Note the minimal overlap between the $3\sigma$ contour and $w=-1$ in the optimum case. Parameter combinations which result in a quintom scenario are shaded in blue. These regions rotate along with the ellipse.}
     \label{fig:guassian contour}
\end{figure*}

\subsection{One Dimensional Tensions \label{subsection: 1-d tension}}

Using the optimal redshift, we formulate the following 1-dimensional null hypothesis: 
\begin{equation} 
\mathcal{H}_0:\qquad w_{\rm opt}=-1. 
\label{eq:null}
\end{equation} 
Under this hypothesis, the equation of state evaluated at the optimal scale factor is assumed to coincide with the cosmological-constant value. The statistical significance of the departure from this null hypothesis may be computed as as follows.

In \cref{subsection: optimum}, $\Delta\chi^2_{\rm1D}=-\frac{(w_*+1)^2}{\langle\delta w_*^2\rangle}$ is used to find the optimum and predict the tension at $a_{\rm opt}$ using sampled CPL parameters alone. Ultimately, \cref{chi opt} is used to derive the optimal $\Delta\chi^2$ and tension using the 1-dimensional relation $|\Delta\chi_{\rm 1D}^2|=\sigma^2$. Values analytically computed from CPL parameters in this way are referred to as "Derived" values.

However, one may also sample parameters using CPL$_*$ where $a_*=a_{\rm opt}$ to obtain a sampled $w_{\rm opt}$ from the mean of the posterior distribution. The variance of the posterior $\langle\delta w_{\rm opt}^2\rangle$ can be used to find
\begin{equation}
    \sigma_{\rm opt} = \frac{|w_{\rm opt}+1|}{\sqrt{\langle\delta w_{\rm opt}^2\rangle}}.
    \label{guassian w tension}
\end{equation}
Values computed in this way are referred to as "Sampled".

\begin{table*}

\begin{tabular}{|c| c|}
\hline
Parameter & Prior \\
\hline
$H_0$ & $\mathcal{U}[20,100]$ \\
$\Omega_bh^2$ & $\mathcal{N}[0.02242,0.00014]$ \\
$\Omega_ch^2$ & $\mathcal{N}[0.011933,0.00091]$ \\
$\log (10^{10}A_s)$ & $\mathcal{U}[1.61,3.91]$ \\
$n_s$ & $\mathcal{U}[0.8,1.2]$ \\
$\tau$ & $\mathcal{U}[0.01,0.1]$ \\
$w_0(w_*)$ & $\mathcal{U}[-3,1]$ \\
$w_a$ & $\mathcal{U}[-3,3]$ \\

\hline
\end{tabular}
\caption{\label{tab:prior} Priors}
\end{table*}

Despite getting $w_{\rm opt}$ and it's variance directly from sampling, when computing tension from the variances and covariances of the posterior alone, important information about the full distribution is often lost. In one dimension, we effectively are approximating it using a Gaussian distribution centered at the mean of the full distribution with a standard deviation equivalent to the square root of the variance. While this approximation is still instructive and helpful for analytically deriving the optimum, it can become poor for computing the true tension of the distribution when significant non-Gaussianities are introduced.

As an alternative method, one may use the 1-D distribution of $w_*$, marginalizing over all other parameters, and find the sample density at $w_*=-1$. Integration may be performed over all $w_*$ parameter space with a lower density (less likely). The fraction of the area with a lower density verses the total area gives the p-value, $p$. One may then convert between $1-p$ and $\sigma$ using
\begin{equation}
    1-p={\rm erf}\bigg(\frac{\sigma}{\sqrt2}\bigg).
    \label{accurate w tension}
\end{equation}

The sampled tensions in \cref{tab:significance} are computed using \cref{guassian w tension} while \cref{tab:alt significance} in appendix \ref{appendix: alt tensions} lists sampled tensions computed using the alternative method and \cref{accurate w tension}.

\section{Data \label{section:data}}

\begin{itemize}

\item \textbf{Baryon Acoustic Oscillations (BAO):} 
We include BAO distance measurements from the DESI Data Release 2 (DR2) \cite{DESI:2025zgx}. This compilation spans a wide redshift range and incorporates multiple tracers, providing strong geometric constraints on the expansion history. 
In addition, we use the Dark Energy Survey Year 6 (DESY6) BAO measurement \cite{weaverdyck2026darkenergysurveyyear}, constructed from a sample optimized for BAO analyses. To ensure statistical independence between datasets, we adopt the DESI-independent DESY6 BAO measurement with overlap removed when combining with DESI data. 

Through this work DESI DR2 will be denoted as DESI2 and DESY6 BAO will be denoted as D6.

\item \textbf{Type Ia Supernovae (SNeIa):} 
We consider several supernova compilations, analyzing one dataset at a time within each combination. The DES Year 5 (DESY5) sample \cite{descollaboration2025darkenergysurveycosmology} consists of photometrically identified SNeIa calibrated with external low-redshift anchors. The Union3 compilation \cite{2016AAS...22713918R} provides an alternative dataset with different light-curve fitting and calibration strategies. We also include the Pantheon+ sample \cite{2022ApJ...938..113S}, which contains a large collection of spectroscopically confirmed SNeIa spanning a broad redshift range.

In addition, we incorporate recently updated or recalibrated versions of these datasets. The DES-Dovekie sample \cite{popovic2026darkenergysurveysupernova} presents an improved calibration of the DESY5 dataset with updated cross-survey consistency. The Union3.1 dataset \cite{hoyt2026union31selfconsistentmeasurementshost} applies a unified treatment of host-galaxy properties across the compilation, and the same methodology is used to produce an updated Pantheon+ variant (denoted PP\_Hoyt in this work).

\begin{figure*}
    \centering
    \begin{subfigure}[b]{0.47\textwidth}
        \centering
        \includegraphics[width=\textwidth]{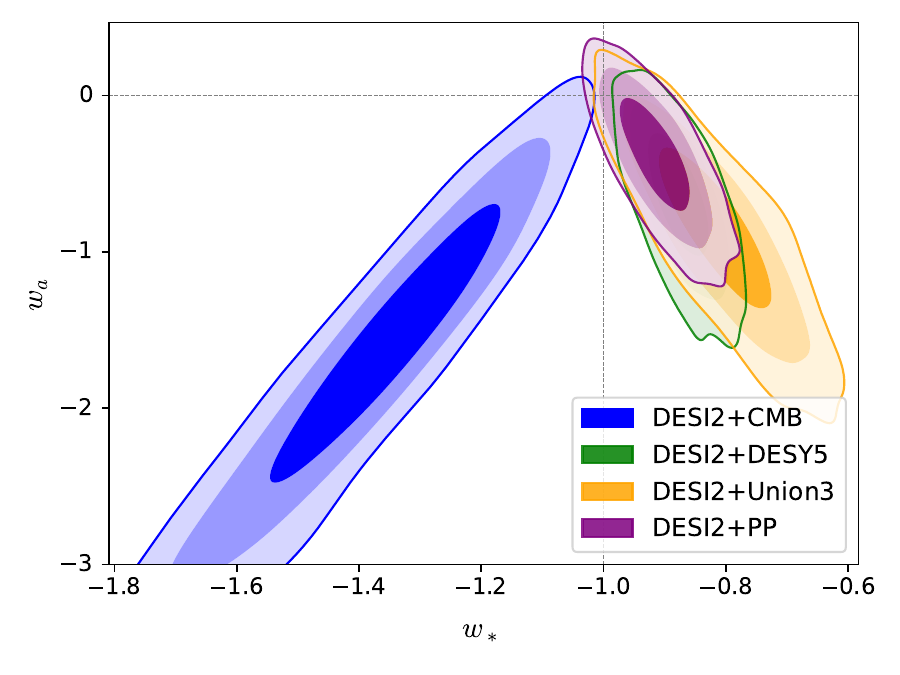}
        \caption{Combinations including previous SN datasets}
    \end{subfigure}
    \begin{subfigure}[b]{0.47\textwidth}
        \centering
        \includegraphics[width=\textwidth]{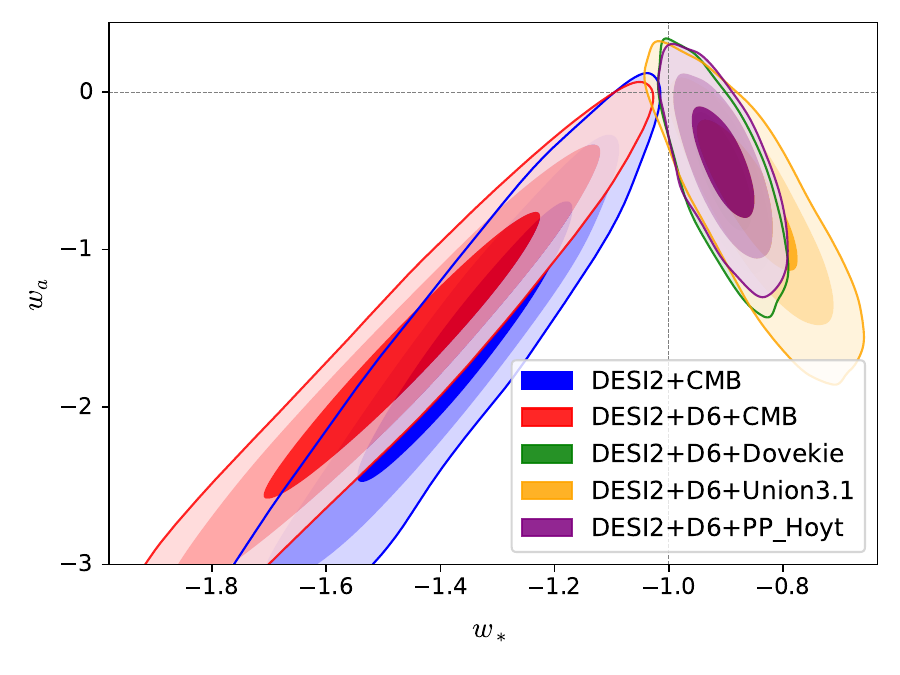}
        \caption{Combinations including re-calibrated SN datasets}
    \end{subfigure}
    \hfill
        \caption{BAO + CMB vs BAO + SN. Shows 1, 2, and 3$\sigma$ contours \footnote{corresponds to 68.3\%, 95.5\%, and 99.7 \% probability respectively.}. The $w_*$ parameter corresponds to $w(z_*)$ where $z_*=z_{\rm opt}$ (or $z_*=0$ if $z_{\rm opt} < 0$) which varies for each dataset. These confidence contour plots are not to be confused with the usual $w_0,w_a$ contour plots which provide estimates of the same value of $w$ (the present value, $w_0$), and where tensions between datasets indicate disagreement on that single value. Rather, tensions between datasets in these plots indicate different estimates of $w$ values for different redshifts. Notably, BAO + CMB distributions show greater than $3\sigma$ preference for $w<-1$. However, there is some overlap between BAO + SN datasets and the $w=-1$ line indicating less than $3\sigma$ tension for $w>-1$ with BAO + SN alone. Note that re-calibrated datasets are in much tighter agreement than the previous ones.}
        \label{fig:BAO+CMB v BAO+SN}
\end{figure*}

Throughout this work, previous supernova datasets DES Year 5, Union3 and Pantheon+ will be denoted as DESY5, Union3, and PP respectively. Re-calibrated datasets will be denoted as Dovekie, Union3\_1, and PP\_Hoyt.

\item \textbf{Cosmic Microwave Background (CMB) and Lensing:} 
We include constraints from the Planck Data Release 3 \cite{2020}, incorporating temperature and polarization measurements of the CMB anisotropies combined with CMB lensing measurements from the Atacama Cosmology Telescope (ACT) \cite{Madhavacheril_2024}. This extended dataset provides enhanced sensitivity to the growth of structure and late-time cosmic evolution. We also added temperature and polarization data from the South Pole Telescope (SPT) for further constraining power \cite{7wt3-9v2y}.

Throughout this work, Planck CMB data along with ACT lensing will be denoted as CMB. Power spectra data from SPT will be denoted as SPT.

\end{itemize}

All parameter estimation is preformed using Metropolis Hasting Monte Carlo Markov Chain (MCMC) sampling via the cosmological inference code Cobaya. CAMB was utilized as the cosmological theory code. For BAO, Supernova, and CMB combined datasets the following parameters are sampled: $[H_0,\Omega_bh^2,\Omega_ch^2,A_s,n_s,\tau,w_0 ( w_*),w_a]$. For BAO and Supernova combined datasets, the perturbation parameters are held fixed at the standard Planck values: $A_s=2.10\cdot 10^{-9}, n_s=0.965, \tau=0.054$. For all datasets, neutrino parameters were held constant at $m_\nu=0.06$ and $n_\nu=3.044$. Choices for priors on these parameters are given in \cref{tab:prior}. 

To obtain the final results, parameter estimation is first performed using the CPL parameterization. From the mean of the sampled $w_0$, $w_a$ as well as their covariance matrix, the scale factor of the optimum $a_{\rm opt}$ and derived $w_{\rm opt} = w(a_{\rm opt})$ are computed using \cref{optimum definition}. A second parameter estimation is then performed using a CPL$_*$ parameterization with $a_*$ chosen to be $a_{\rm opt}$. The mean of the sampled $w_*$ is recorded as the sampled $w_{\rm opt}$. Sampled tensions are computed from this value and the variance of $w_{\rm opt}$ from the posterior distribution. Alternative values for sampled tensions are also computed using the method described in \cref{subsection: 1-d tension}. Derived tensions are computed from the derived $w_{\rm opt}$ and $\langle\delta w_{\rm opt}^2\rangle$ which assumes a Gaussian approximation of the distribution. Tensions with $\rm \Lambda CDM$ are computed via $\Delta \chi^2$ between the CPL parameter estimation and a subsequent $\rm \Lambda CDM$ parameter estimation using minimization from iminuit.

\section{Results \label{section:results}}

Analysis of individual dataset combinations is detailed in \cref{tab:parameter constraints} and \cref{tab:significance}. CPL and CPL$_*$ parameter estimates can be found in \cref{tab:parameter constraints} for each dataset combination used. Tensions between $w_{\rm opt}$ and -1 as well as tensions with $\rm \Lambda CDM$ can be found in \cref{tab:significance}.

Analyses between different dataset combinations are explored in the figures. Dataset combinations are grouped into three types, BAO + CMB, BAO + SN, and BAO + SN + CMB. BAO + CMB is used in all plots as this is the only combination type where $w_{\rm opt}<-1$ and therefore can be used as a type 2 dataset (i.e. the equation of state is below the phantom-crossing line). BAO + SN have $w_{\rm opt}>-1$ and serve as type 1 combinations (above the phantom crossing line). The triplet combinations, BAO + SN + CMB, also have $w_{\rm opt}>-1$ and are used as type 1 combinations. However, these triplet combinations consistently have $a_{\rm opt}>1$ which implies that the optimal values of $w_*$ are located in the future and have not yet been reached. As noted at the end of \cref{subsection: optimum}, in these cases, $w_0$ (instead of $w_{\rm opt}$) is taken as our $w_1$ and all plots showing BAO + SN + CMB at the optimum instead display their original CPL posterior likelihoods. Since $w_0 > -1$, these datasets can still serve as type 1 combinations giving a $w_1>-1$. Finally, we note that the interesting result here is the significance levels above 3-sigma inside the phantom regime for the equation of state, as we discuss below.

\begin{figure*} 
     \centering
     \begin{subfigure}[b]{0.47\textwidth}
         \centering
         \includegraphics[width=\textwidth]{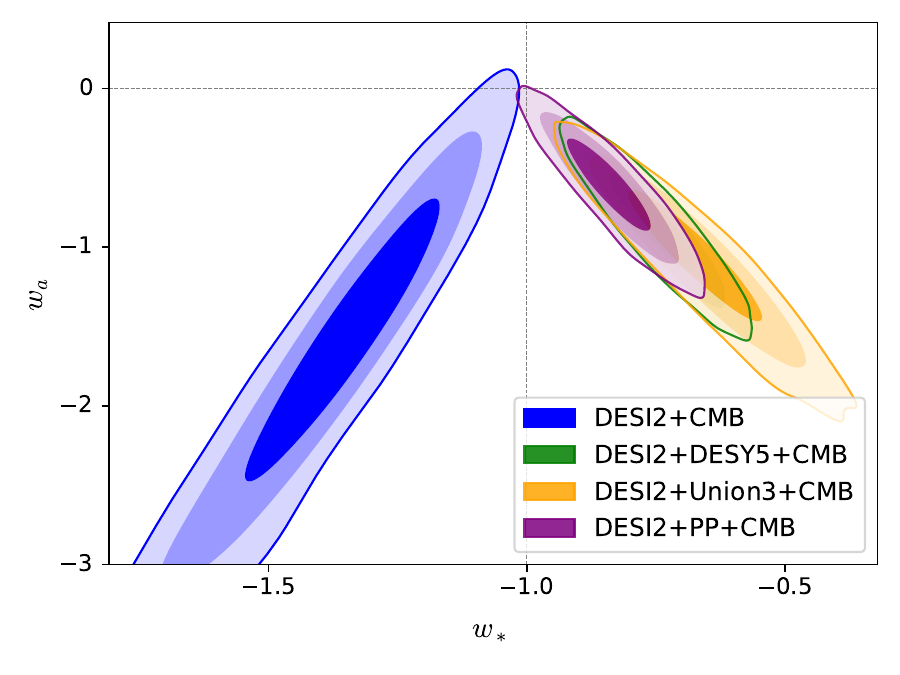}
         \caption{Previous SN Datasets}
     \end{subfigure}
     \begin{subfigure}[b]{0.47\textwidth}
         \centering
         \includegraphics[width=\textwidth]{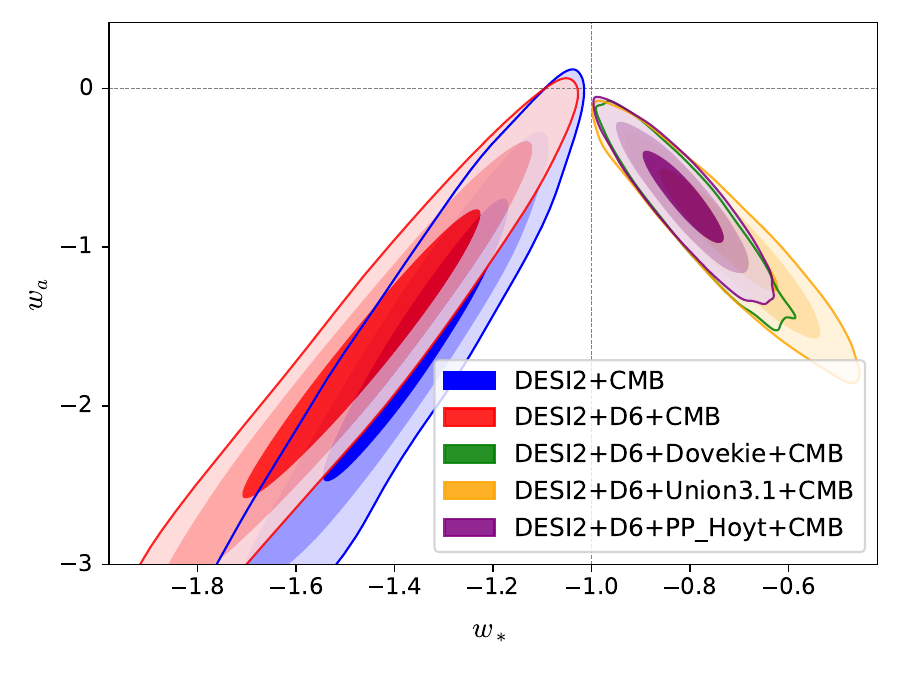}
         \caption{Re-Calibrated SN Datasets}
     \end{subfigure}
     \hfill
        \caption{BAO + CMB vs BAO + SN + CMB. Shows 1, 2, and 3$\sigma$ contours. With the exception of DESI2 + PP + CMB, there is no overlap with $w=-1$ for BAO + SN + CMB combinations, with a notable gap present. This indicates stronger than $3\sigma$ tension between the $w_{\rm opt}$ estimates and the $w=-1$ crossing line in addition to the over $3\sigma$ preference for $w<-1$ indicated by BAO + CMB. The gap between distribution contours indicates that the two dataset combinations constrain the equation of state to be on two different sides of the phantom line $w=-1$. Note that re-calibrated datasets are in much tighter agreement than previous ones.}
    \label{fig:BAO+CMB v BAO+SN+CMB}
\end{figure*}

\begin{figure*} 
     \centering
     \begin{subfigure}[b]{0.47\textwidth}
         \centering
         \includegraphics[width=\textwidth]{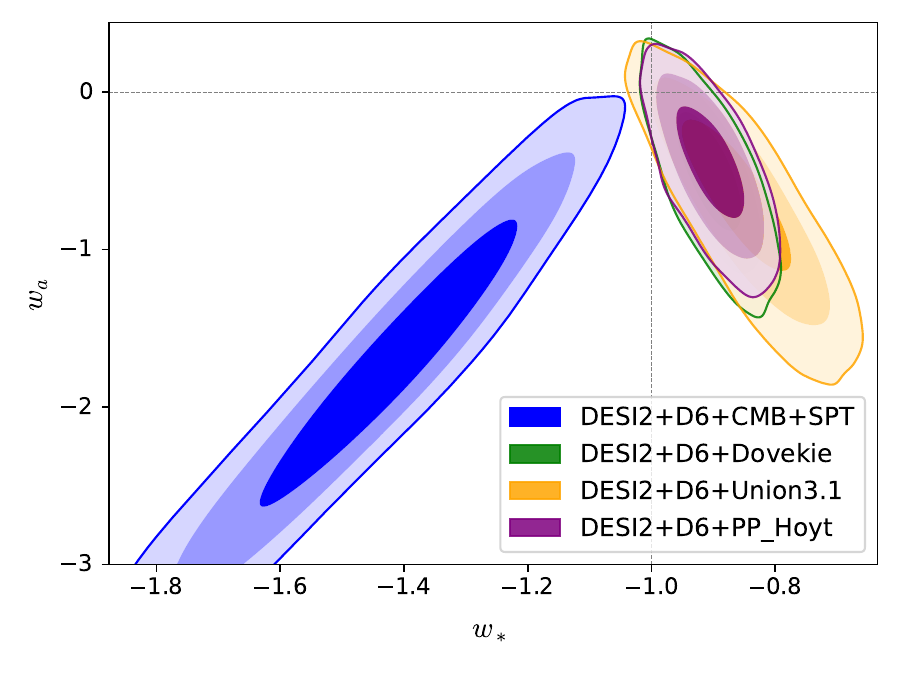}
         \caption{BAO + SN}
         \label{subfig:Cadillac A}
     \end{subfigure}
     \begin{subfigure}[b]{0.47\textwidth}
         \centering
         \includegraphics[width=\textwidth]{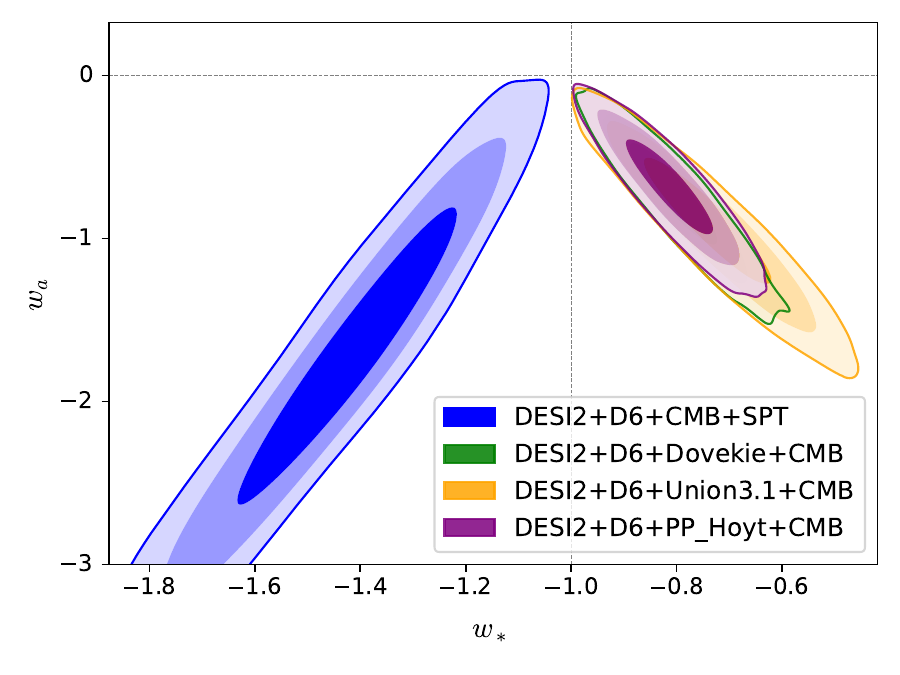}
         \caption{BAO + SN + CMB}
         \label{subfig:Cadillac B}
     \end{subfigure}
     \hfill
        \caption{Comparisons to DESI2 + D6 + CMB + SPT (Re-Calibrated SN Datasets). Shows 1, 2, and 3$\sigma$ contours. Both plots use re-calibrated supernova data. The left plot compares DESI2 + D6 + CMB + SPT with BAO + SN dataset combinations while the right compares with BAO + SN + CMB. When including SPT, the BAO + CMB distribution moves further away from $w=-1$, creating a slight gap between BAO + SN. Comparing to BAO + SN + CMB shows largest tension on both sides as well as the largest gap between distributions.  Again, the gap between distribution contours indicates that the two dataset combinations constrain the equation of state to be on two different sides of the phantom line $w=-1$. There is a tension of over $3\sigma$ on both sides with $-1$. There is also a tension of over $3\sigma$ between the distributions of the two dataset combinations, indicating that the EOS lies on different sides of the phantom line $w=-1$.}
    \label{fig:Cadillac}
\end{figure*}

In the first section of \cref{tab:parameter constraints} and \cref{tab:significance}, the parameter estimates and tensions for the type 2 dataset combinations (BAO + CMB) can be found. All $w_{\rm opt}$ in this section are below -1, falling between -1.358 and -1.464 and the sampled $\sigma_{\rm opt}$ tensions are found to be between 3.01 and 3.22 with the high end value requiring the addition of SPT data in the DESI2 + D6 + CMB + SPT combination\footnote{Unless otherwise stated, sampled $\sigma_{\rm opt}$ values written out in the results section are computed using \cref{guassian w tension}}. Note that these tensions indicate values of the equation of state parameter that are in the phantom regime.

Secondly, the type 1 dataset triplet combinations (BAO + SN + CMB), when including the most recent re-calibrated supernova data (Dovekie, Union3\_1, PP\_Hoyt), give consistent tension results for $\sigma_{\rm opt}$ between 3.30 and 3.55. Though the true $w_{\rm opt}$ values widely vary and occur at future redshifts, the $w_0$ values used for the tensions are also consistent and lie between $-0.737$ and $-0.812$ showing tight agreement between the re-calibrated datasets. BAO + SN (with re-calibrated SN) also have similar tensions with each other, between $2.47$ and $2.76$ and $w_{\rm opt}$ values between $-0.859$ and $-0.906$.

 Combining the datasets which give the strongest tension yields a $3.55\sigma$ tension above $w=-1$ and, most noteworthy, $3.22\sigma$ tension below.

Predictions for the redshift of the phantom crossing can somewhat vary between dataset combinations. The BAO + CMB combinations consistently predict $z_c\approx0.50$, while triplet combinations prefer a later value of $z_c$ between 0.35 and 0.45. BAO + SN combinations yield a $z_c$ between 0.44 and 0.51, in between that of the previous two.

When analyzing the contour plots in \cref{fig:BAO+CMB v BAO+SN} through \cref{fig:Cadillac}, it is clear that the 3 sigma contours on all type 2 dataset combinations are not close to intersecting the $w=-1$ line. At first, this seems not to align with the data in \cref{tab:significance} that predicts $\sigma_{\rm opt}$ is not much greater than $3$. This discrepancy is due to a significant non-Gaussianity in the posterior distribution which causes the Gaussian estimations of standard deviation made in \cref{tab:significance} to underestimate the true tension. An example of the Gaussian estimate of the posterior distribution can be seen in \cref{fig:guassian contour} which plots the derived ellipse contours for DESI2 + D6 + CMB + SPT. Here, the 3-sigma contour barely intersects the $w=-1$ line, which, when marginalized over $w_a$ produces a $\sigma_{\rm opt} \sim 3.2$. In these cases, it is informative to view the non-Gaussian tensions computed using the alternative method detailed in \cref{subsection: 1-d tension}. Tension metrics computed using this method can be found in the \cref{tab:alt significance} located in the appendix. When considering these values the maximum 1-D tensions for BAO + CMB dataset combinations lay between $3.85\sigma$ and $4.17\sigma$, a notable increase.

The most significant results can be seen when considering BAO + SN + CMB which all do not intersect $w=-1$ at 3-sigma (for re-calibrated SN). This, in combination with the BAO + CMB datasets, leaves a notable gap between the 3-sigma contours of the two probability distributions. The strongest case of this is shown in \cref{subfig:Cadillac B} which compares the re-calibrated BAO + SN + CMB combinations with DESI2 + D6 + CMB + SPT, the most robust BAO + CMB combination, showing the largest gap obtained. Such a notable gap indicates that one dataset combination prefers $w$ firmly in the $w<-1$ phantom regime while the other prefers $w>-1$.

Meanwhile, the 3-sigma contours of all BAO + SN datasets intersect $w=-1$ and overlap with the 3-sigma contours of most BAO + CMB dataset combinations. DESI2 + D6 + CMB + SPT being a notable exception, having zero overlap with any BAO + SN combination.

Finally, for completeness, dataset combinations involving the previous supernova datasets (DESY5, Union3, PP) were also included. In these dataset combinations, only DESI2 is included in the BAO. Significances with these datasets are typically stronger than with the re-calibrated with the exception of combinations involving PP which have notably lower tensions. As these datasets are largely being replaced by the new re-calibrated ones, these results are not central to our discussion but are provided for completeness and can still be viewed in \cref{tab:parameter constraints} and \cref{tab:significance} as well as \cref{fig:BAO+CMB v BAO+SN} and \cref{fig:BAO+CMB v BAO+SN+CMB}.

\begin{table*}
\begin{threeparttable}
\centering
\footnotesize
\setlength{\tabcolsep}{4pt}
\renewcommand{\arraystretch}{1.2}

\begin{tabular}{@{}lcccccc@{}}
\hline\hline
Dataset & $w_0$ & $w_a$ & $a_{\rm opt}\,(z_{\rm opt})$ & $a_c\,(z_c)$ & Deriv.\ $w_{\rm opt}$ & Samp.\ $w_{\rm opt}$ \\

\hline
\multicolumn{7}{c}{\textit{BAO + CMB}} \\
\hline
DESI2 + CMB & $-0.456\pm0.203$ & $-1.624\pm0.558$ & 0.428 (1.336) & 0.665 (0.504) & $-1.386\pm0.126$ & $-1.385\pm0.128$ \\
DESI2 + D6 + CMB & $-0.445\pm0.205$ & $-1.654\pm0.566$ & 0.388 (1.577) & 0.665 (0.504) & $-1.458\pm0.151$ & $-1.464\pm0.151$ \\
DESI2 + D6 + CMB + SPT & $-0.421\pm0.200$ & $-1.743\pm0.551$ & 0.423 (1.364) & 0.668 (0.497) & $-1.427\pm0.128$ & $-1.420\pm0.130$ \\

\hline
\multicolumn{7}{c}{\textit{BAO + SN + CMB}} \\
\hline
DESI2 + D6 + Dovekie + CMB & $-0.803\pm0.055$ & $-0.737\pm0.207$ & 1.038 (-0.036) & 0.733 (0.364) & $-0.599\pm0.11$ & $-0.803\pm0.055^{a}$ \\
DESI2 + D6 + Union3.1 + CMB & $-0.737\pm0.080$ & $-0.909\pm0.258$ & 10.98 (-0.909) & 0.71 (0.408) & $8.34\pm2.65$ & $-0.737\pm0.080^{a}$ \\
DESI2 + D6 + PP\_Hoyt + CMB & $-0.812\pm0.054$ & $-0.687\pm0.193$ & 1.416 (-0.294) & 0.727 (0.376) & $-0.526\pm0.131$ & $-0.812\pm0.054^{a}$ \\

DESI2 + DESY5 + CMB & $-0.753\pm0.056$ & $-0.851\pm0.206$ & 1.03 (-0.029) & 0.709 (0.410) & $-0.727\pm0.061$ & $-0.753\pm0.056^{a}$ \\
DESI2 + Union3 + CMB & $-0.671\pm0.084$ & $-1.065\pm0.272$ & 1.338 (-0.253) & 0.691 (0.447) & $-0.311\pm0.173$ & $-0.671\pm0.084^{a}$ \\
DESI2 + PP + CMB & $-0.838\pm0.053$ & $-0.616\pm0.192$ & 2.052 (-0.513) & 0.738 (0.355) & $-0.19\pm0.251$ & $-0.838\pm0.053^{a}$ \\

\hline
\multicolumn{7}{c}{\textit{BAO + SN}} \\
\hline
DESI2 + DESY5 & $-0.787\pm0.061$ & $-0.649\pm0.254$ & 0.865 (0.156) & 0.672 (0.488) & $-0.875\pm0.032$ & $-0.877\pm0.032$ \\
DESI2 + Union3 & $-0.717\pm0.094$ & $-0.841\pm0.326$ & 0.881 (0.135) & 0.664 (0.506) & $-0.818\pm0.059$ & $-0.816\pm0.061$ \\
DESI2 + PP & $-0.878\pm0.058$ & $-0.38\pm0.234$ & 0.898 (0.114) & 0.678 (0.475) & $-0.917\pm0.038$ & $-0.914\pm0.037$ \\

DESI2 + D6 + Dovekie & $-0.843\pm0.062$ & $-0.509\pm0.258$ & 0.78 (0.282) & 0.692 (0.445) & $-0.905\pm0.035$ & $-0.906\pm0.034$ \\
DESI2 + D6 + Union3.1 & $-0.779\pm0.087$ & $-0.685\pm0.317$ & 0.887 (0.127) & 0.678 (0.475) & $-0.857\pm0.055$ & $-0.859\pm0.057$ \\
DESI2 + D6 + PP\_Hoyt & $-0.852\pm0.061$ & $-0.449\pm0.244$ & 0.880 (0.136) & 0.670 (0.493) & $-0.905\pm0.036$ & $-0.904\pm0.035$ \\

\hline\hline
\end{tabular}

\caption{Parameter constraints for CPL and optimum. Derived values of $w_{\rm opt}$ are obtained by evaluating $w(a)$ using the CPL parameters at $a_{\rm opt}$, while sampled values correspond to the mean $w_*$ from the CPL$_*$ parameterization centered at $a_*=a_{\rm opt}$.}
\label{tab:parameter constraints}
\begin{tablenotes}
\item[a] For $a_{\rm opt}>1$, the sampled value corresponds to $w_0$ instead of $w_{\rm opt}$.
\end{tablenotes}

\end{threeparttable}
\end{table*}

\begin{table*}
\begin{threeparttable}
\centering
\footnotesize
\setlength{\tabcolsep}{4pt}
\renewcommand{\arraystretch}{1.2}

\begin{tabular}{@{}lccccccc@{}}
\hline\hline
Dataset & Deriv.\ $\sigma_{\rm opt}$ & Samp.\ $\Delta \chi^2_{\rm opt}$ & Samp.\ $\sigma_{\rm opt}$ & Samp.\ $\Delta \chi^2$ & $\Lambda$CDM& $z_{\rm opt}$ & Samp.\ $w_{\rm opt}$ \\
& & & & & tension &  &  \\
\hline

\multicolumn{8}{c}{\textit{BAO + CMB}} \\
\hline
DESI2 + CMB & $\mathbf{3.06\ (< -1)}$ & -9.05  & $\mathbf{3.01\ (< -1)}$ & -11.75 & 2.99 & 1.336 & $-1.385 \pm 0.128$\\
DESI2 + D6 + CMB & $\mathbf{3.03\ (< -1)}$ & -9.46  & $\mathbf{3.08\ (< -1)}$ & -11.81 & 3.00 & 1.577 & $-1.464 \pm 0.151$\\
DESI2 + D6 + CMB + SPT & $\mathbf{3.32\ (< -1)}$ & -10.38  & $\mathbf{3.22\ (< -1)}$ & -15.21 & 3.48 & 1.364 & $-1.420 \pm 0.130$\\

\hline
\multicolumn{8}{c}{\textit{BAO + SN + CMB}} \\
\hline
DESI2 + D6 + Dovekie + CMB & $\mathbf{3.63\ (> -1)}$ & -12.63$^{a}$  & $\mathbf{3.55\ (> -1)}^{a}$ & -13.58 & 3.26 & - & $-0.803 \pm 0.055^{a}$\\
DESI2 + D6 + Union3.1 + CMB & $\mathbf{3.52\ (> -1)}$ & -10.86$^{a}$ & $\mathbf{3.30\ (> -1)}^{a}$ & -13.67 & 3.27 & - & $-0.737 \pm 0.080^{a}$\\
DESI2 + D6 + PP\_Hoyt + CMB & $\mathbf{3.62\ (> -1)}$ & -12.13$^{a}$ & $\mathbf{3.48\ (> -1)}^{a}$ & -13.19 & 3.20 & - & $-0.812 \pm 0.054^{a}$\\

DESI2 + DESY5 + CMB & 4.45 ($> -1$) & -19.78$^{a}$  & 4.45 ($> -1$)$^{a}$ & -19.60 & 4.03 & - & $-0.753 \pm 0.056^{a}$\\
DESI2 + Union3 + CMB & 3.97 ($> -1$) & -15.22$^{a}$  & 3.90 ($> -1$)$^{a}$ & -16.90 & 3.70 & - & $-0.671 \pm 0.084^{a}$\\
DESI2 + PP + CMB & 3.23 ($> -1$) & -9.16$^{a}$ & 3.03 ($> -1$)$^{a}$ & -10.33 & 2.76 & - & $-0.838 \pm 0.053^{a}$\\

\hline
\multicolumn{8}{c}{\textit{BAO + SN}} \\
\hline
DESI2 + D6 + Dovekie & 2.73 ($> -1$) & -7.62  & 2.76 ($> -1$) & -7.36 & 2.24 & 0.282 & $-0.906 \pm 0.034$\\
DESI2 + D6 + Union3.1 & 2.61 ($> -1$) & -6.12 & 2.47 ($> -1$) & -6.45 & 2.06 & 0.127 & $-0.859 \pm 0.057$\\
DESI2 + D6 + PP\_Hoyt & 2.60 ($> -1$) & -7.50 & 2.74 ($> -1$) & -7.41 & 2.25 & 0.136 & $-0.905 \pm 0.035$\\

DESI2 + DESY5 & 3.86 ($> -1$) & -14.42 & 3.80 ($> -1$) & -13.60 & 3.26 & 0.156 & $-0.877 \pm 0.032$\\
DESI2 + Union3 & 3.10 ($> -1$) & -9.18 & 3.03 ($> -1$) & -10.08 & 2.72 & 0.135 & $-0.816 \pm 0.061$\\
DESI2 + PP & 2.20 ($> -1$) & -5.22 & 2.28 ($> -1$) & -4.92 & 1.72 & 0.114 & $-0.914 \pm 0.037$ \\

\hline\hline
\end{tabular}

\caption{Chi-squared and tension metrics.
Derived quantities are computed using Gaussian approximations from CPL parameter estimations, while sampled quantities are obtained from posterior distributions from CPL$_*$ centered at $a_*=a_{\rm opt}$. Values in boldface correspond to the key combinations used in our discussion, all providing significance levels in excess of $3\sigma$ above and (interestingly) below the $-1$ phantom-crossing line, using the most recently re-calibrated datasets when SN are included.}
\label{tab:significance}

\begin{tablenotes}
\item[a] For $a_{\rm opt}>1$, the sampled value corresponds to $w_0$ instead of $w_{\rm opt}$.
\end{tablenotes}

\end{threeparttable}
\end{table*}

\section{Conclusion \label{section:conclusion}}

In paper-I \cite{paper-I}, we introduced the concept of the optimal redshift, which aims to simultaneously minimize the error on the equation of state parameter $w$ and maximize the distance between $w$ and the phantom crossing line for a given dataset combination. In this work (paper-II), we applied the corresponding formalism to datasets and analyzed the evidence for a phantom crossing given the CPL parameterization. Using a variety of different dataset combinations including newly re-calibrated supernova data we were able to compute the optimum $a_{\rm opt}$ of each dataset combination and perform a sampling using the CPL$_*$ parameterization with $a_*$ chosen to be $a_{\rm opt}$. We then compared the value of $w(a_{\rm opt})$ to the null hypothesis $\mathcal{H}_0: w_{\rm opt}=-1$. By using the posterior distributions of these chains, the tension above and below the crossing line $w=-1$ could be analyzed and specifically estimated.

From \cref{section:results}, it was seen that the majority of dataset combinations including re-calibrated supernovae have maximal 1-dimensional tensions of at least $3\sigma$ with $w=-1$ on both sides. Dataset combinations of the type BAO + CMB are particularly effective at  constraining $w(a)$ before the crossing and dataset combinations of the type BAO + SN + CMB are particularly effective at constraining $w(a)$ after the crossing. It is well known that present day values of $w(a)$ lie above $-1$. We find significance levels that reach $3.22\sigma$ before the crossing point (i.e. in the phantom regime) and $3.55\sigma$ after the crossing. The new result here is that we can now ascribe a significance of over $3\sigma$ in the phantom region, thereby strongly constraining $w(z)$ on both sides of $-1$ at different redshifts, providing evidence for a crossing at some point between these redshifts within the CPL framework. 

Additionally, when considering the $\rm \Lambda CDM$ or the $w\rm CDM$ models (i.e. constant EOS), differing $w$ values for differing redshifts should not be expected. The value of $w(z)$ should remain constant no matter the scale factor being sampled. Such a large disagreement (over $3\sigma$) between the $w(z)$ values estimated from BAO + CMB at $z_{\rm opt} \sim 1.4$ datasets and $w_0$ of BAO + SN + CMB at $z=0$ provides even further evidence for an evolving equation of state and significant deviations from $\rm \Lambda CDM$.

It is worth noting that the goal of this work was not to discuss whether the crossing is an effective or intrinsic one, but rather to find a framework to enhance its evidence in the past before the crossing point, given the CPL parameterization (an its generalization).  Using the optimum redshift point, one can better quantitatively evaluate the evidence provided for a phantom crossing when using CPL on the most recent host of cosmological data.  Indeed, it remains to be explored what the true micro-physical nature of these results are. 

Finally, as future surveys deliver increasingly precise measurements, the optimal-redshift framework offers a promising avenue for maximizing the constraining power of cosmological observations on the nature and evolution of dark energy, particularly with regard to detecting deviations from $w=-1$ and characterizing possible phantom-crossing transitions. Such applications are especially timely in the era of DESI, Rubin LSST, Euclid, the Nancy Grace Roman Space Telescope, the Simons Observatory and other next-generation cosmological surveys.

\section{Acknowledgements}
We thank David Shlivko and Kushal Lodha for providing useful comments on the manuscript. The authors acknowledge that Gemini was used to help with writing scripts for the plots in this paper. MI acknowledges that this material is based upon work supported in part by the Department of Energy, Office of Science, under Award Number DE-SC0022184 and also in part by the U.S. National Science Foundation under grant AST2327245.
\clearpage
\appendix

\section{Alternative Tension Metrics} \label{appendix: alt tensions}

Tension metrics for the datasets tested in this work may also be computed using the alternative method detailed in \cref{subsection: 1-d tension}. This method integrates over the entire posterior distribution to compute tensions from the p-value of the likelihood using \cref{accurate w tension}. These results are listed in \cref{tab:alt significance}.

\begin{table*}[hp]
\begin{threeparttable}
\centering
\footnotesize
\setlength{\tabcolsep}{4pt}
\renewcommand{\arraystretch}{1.2}

\begin{tabular}{@{}lccc@{}}
\hline\hline
Dataset & Deriv.\ $\sigma_{\rm opt}$ & Samp.\ $\Delta \chi^2_{\rm opt}$ & Samp.\ $\sigma_{\rm opt}$ \\

\hline

\multicolumn{4}{c}{\textit{BAO + CMB}} \\
\hline
DESI2 + CMB & $\mathbf{3.06\ (< -1)}$ & 15.05 & $\mathbf{3.88\ (< -1)}$\\
DESI2 + D6 + CMB & $\mathbf{3.03\ (< -1)}$ & 14.80 & $\mathbf{3.85\ (< -1)}$\\
DESI2 + D6 + CMB + SPT & $\mathbf{3.32\ (< -1)}$ & 17.35 & $\mathbf{4.17\ (< -1)}$\\

\hline
\multicolumn{4}{c}{\textit{BAO + SN + CMB}} \\
\hline
DESI2 + D6 + Dovekie + CMB & $\mathbf{3.63\ (> -1)}$ & 14.33$^{a}$ & $\mathbf{3.79\ (> -1)}^{a}$ \\
DESI2 + D6 + Union3.1 + CMB & $\mathbf{3.52\ (> -1)}$ & 12.01$^{a}$ & $\mathbf{3.47\ (> -1)}^{a}$\\
DESI2 + D6 + PP\_Hoyt + CMB & $\mathbf{3.62\ (> -1)}$ & 13.21$^{a}$ & $\mathbf{3.63\ (> -1)}^{a}$\\

DESI2 + DESY5 + CMB & 4.45 ($> -1$) & 18.50$^{a}$ & 4.30 ($> -1$)$^{a}$\\
DESI2 + Union3 + CMB & 3.97 ($> -1$) & 17.48$^{a}$ & 4.18 ($> -1$)$^{a}$ \\
DESI2 + PP + CMB & 3.23 ($> -1$) & 9.42$^{a}$ & 3.07 ($> -1$)$^{a}$\\

\hline
\multicolumn{4}{c}{\textit{BAO + SN}} \\
\hline
DESI2 + D6 + Dovekie & 2.73 ($> -1$) & 7.75 & 2.78 ($> -1$)\\
DESI2 + D6 + Union3.1 & 2.61 ($> -1$) & 6.68 & 2.58 ($> -1$)\\
DESI2 + D6 + PP\_Hoyt & 2.60 ($> -1$) & 7.75 &  2.78 ($> -1$)\\

DESI2 + DESY5 & 3.86 ($> -1$) & 16.97 & 4.12 ($> -1$) \\
DESI2 + Union3 & 3.10 ($> -1$) & 9.63 & 3.10 ($> -1$)\\
DESI2 + PP & 2.20 ($> -1$) & 5.34 & 2.31 ($> -1$) \\

\hline\hline
\end{tabular}

\caption{Chi-squared and tension metrics using alternative method.}
\label{tab:alt significance}

\begin{tablenotes}
\item[a] $a_{\rm opt}>1$, the sampled value corresponds to $w_0$ instead of $w_{\rm opt}$.
\end{tablenotes}

\end{threeparttable}
\end{table*}

\section{Relation Between Optimal Tension and the Tension with \texorpdfstring{$\rm \Lambda CDM$}{LCDM}} \label{appendix: 2-d tensions}
In this section, we focus on the approximation of tension measures involving two dimensional tensions with the null hypothesis $\rm \Lambda CDM$.

To evaluate the degree to which an alternative hypothesis is favored over a null hypothesis we must consider the likelihood distribution of each model given the data. The likelihood distribution of the alternative model is given as $\mathcal{L}(\theta)$ where $\theta$ is a vector of cosmological parameters. The likelihood will be maximized at some parameter values $\theta_{\rm max}$. The $\Delta \chi^2$ between the alternative and the null hypothesis is given by
\begin{equation}
    \Delta \chi^2=2\log(\mathcal{L}(\theta_{\rm null}))-2\log(\mathcal{L}(\theta_{\rm max})).
\end{equation}

If we preform a multi-variable taylor expansion about $\theta_{\rm max}$ we can approximate the log likelihood around the maximum to second order (a Gaussian approximation) as
\begin{equation}
    \log \mathcal{L}(\theta) \approx \log \mathcal{L}(\theta_{\rm max})+\frac12(\theta-\theta_{\rm max})^T\bm{H}(\theta-\theta_{\rm max})
\end{equation}
where $\bm{H}$ is the Hessian of $\log \mathcal{L}(\theta)$.

By approximating $\log(\mathcal{L}(\theta_{\rm null}))$ in this manner, $\Delta \chi^2$ can be estimated by
\begin{equation}
    \Delta \chi^2\approx\Delta \chi^2_{\rm est} =\delta^T\bm{H}\delta
    \label{general approx chi2}
\end{equation}
where $\delta = \theta_{\rm null} - \theta_{\rm max}$.

We will consider the case where the null hypothesis is the $\rm \Lambda CDM$ model and is formulated as
\begin{equation} 
\mathcal{H}^{\rm \Lambda CDM}_0:\qquad w_{\rm *}=-1 \quad w_a=0, 
\label{eq:lcdm null}
\end{equation} 
while the alternative hypothesis is CPL$_*$. Here we choose to focus specifically on the CPL$_*$ parameters $w_*$ and $w_a$ and, in order to simplify the Hessian, we assume the difference in best fit estimates between models for all other parameters is minimal. In this case $\delta$ is only non-zero for the $w_*,w_a$ elements and contributions from other parameters in the final $\Delta\chi^2$ will not appear. Thus, the Hessian may be expressed as a 2x2 matrix for parameters $w_*,w_a$.
Given that the Hessian is equal to the negative inverse covariance matrix of the parameters, $\bm{H}=-\bm{C}^{-1}$, we may write it as,
\begin{equation}
    \bm{H}= \frac1D\begin{bmatrix}
-\langle \delta w_a^2\rangle & \langle \delta w_* \delta w_a\rangle \\
\langle \delta w_* \delta w_a\rangle & -\langle \delta w_*^2\rangle
\end{bmatrix}
\end{equation}
where $D$ is the determinant of $\bm{H}$ and is given by,
\begin{equation}
    D=\langle \delta w_*^2\rangle\langle \delta w_a^2\rangle-\langle \delta w_* \delta w_a\rangle^2.
\end{equation}

In the general case, a null hypothesis with fixed parameters $w_*=\hat{w}_*$ and $w_a=\hat{w}_a$ will have an estimated $\Delta \chi^2$ of
\begin{equation}
    \Delta \chi_{\rm est}^2=-\frac1D[(\hat{w}_*-w_*)^2\langle \delta w_a^2\rangle-2(\hat{w}_*-w_*)(\hat{w}_a-w_a)\langle \delta w_*\delta w_a\rangle+(\hat{w}_a-w_a)^2\langle \delta w_*^2\rangle].
    \label{general ellipse equation}
\end{equation}
This ellipse equation allows the plotting of estimated n-sigma contours by fixing $\Delta\chi^2_{\rm est}$ as 2.30, 6.18, and 11.83 corresponding to 1, 2 and 3-sigma respectively \footnote{Though sigma can only be properly defined in 1-dimension, these contours correspond to 68.3\%, 95.5\%, and 99.7 \% probability respectively, the same probabilities associated with 1, 2 and 3 sigma tensions in 1-dimension, thus we label them accordingly.}. In practice, these contours can be predicted at any $a_*$ by using CPL parameters $w_0, w_a$ and \cref{w* def} to compute $w_*$. See \cref{fig:guassian contour} for an illustration.

\begin{figure*}
     \centering
     \begin{subfigure}[b]{0.51\textwidth}
         \centering
         \includegraphics[width=\textwidth]{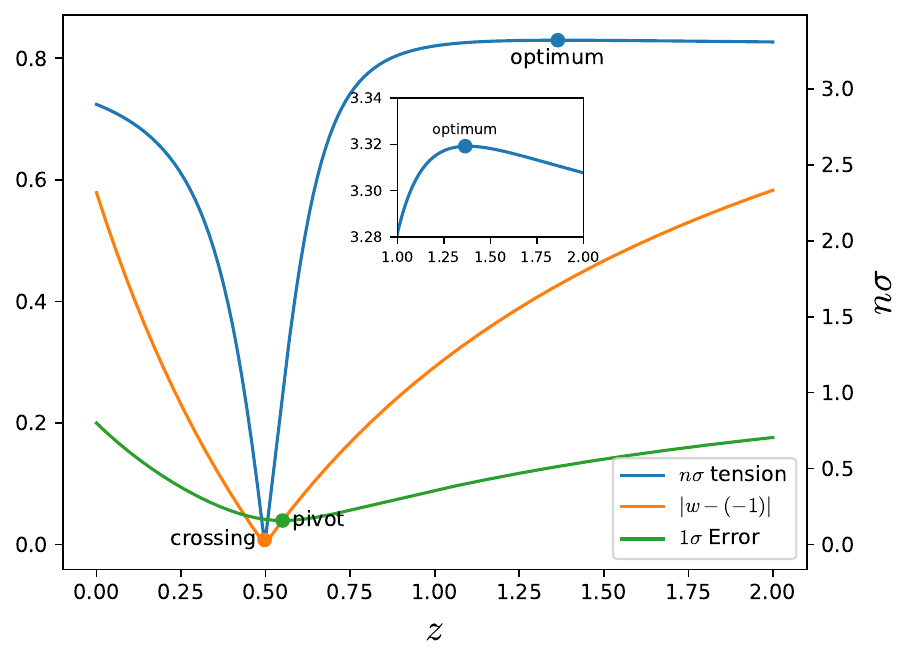}
     \end{subfigure}
     \begin{subfigure}[b]{0.48\textwidth}
         \centering
         \includegraphics[width=\textwidth]{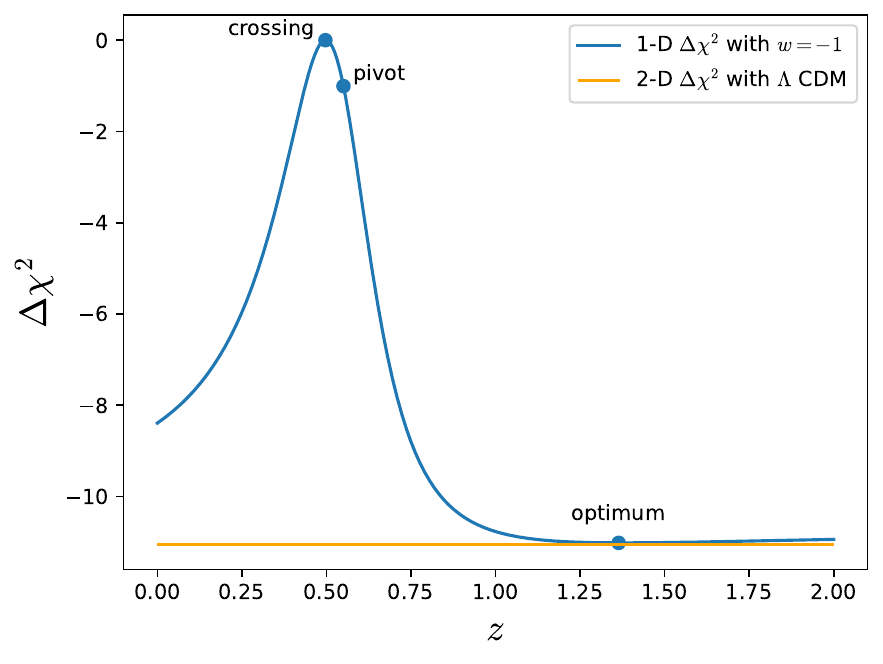}
     \end{subfigure}
     \hfill
        \caption{Estimated Significance at Different $z$. In the left plot, using CPL parameter estimation from the DESI2 + D6 + CMB + SPT dataset combination, both the distance of $w(a)$ from $-1$ as well as the $1\sigma$ error are plotted and use the left y-axis. Taking the ratio of these two gives the n-sigma tension which is plotted using the right y-axis. The smaller inserted plot shows a zoom-in around the optimum, illustrating it's nature as a local maxima. In the right plot, $\Delta\chi_{\rm 1D}^2$ is plotted. Additionally, the derived 2-dimensional $\Delta\chi^2$ calculated using \cref{cpl approx chi2} is plotted, showing it to be constant and equal to $\Delta\chi_{\rm opt}^2$ (-11.05), serving as an upper bound on the 1-dimensional $\chi^2$ (in a Gaussian approximation).}
        \label{fig:estimated chi2}
\end{figure*}

When addressing the $\Delta\chi^2_{\rm est}$ for the null hypothesis of $\rm \Lambda CDM$, $\hat{w}_*=-1$ and $\hat{w}_a=0$. This yields
\begin{equation}
    \Delta \chi_{\rm est}^2=-\frac1D[(w_*+1)^2\langle \delta w_a^2\rangle-2(w_*+1)w_a\langle \delta w_*\delta w_a\rangle+w_a^2\langle \delta w_*^2\rangle].
    \label{cpl approx chi2 a*}
\end{equation}
However, when using equation \cref{w* def}, all dependence on $a_*$ cancels and we are left with a constant value,
\begin{equation}
    \Delta \chi_{\rm est}^2=-\frac1D[(w_0+1)^2\langle \delta w_a^2\rangle-2(w_0+1)w_a\langle \delta w_0\delta w_a\rangle+w_a^2\langle \delta w_0^2\rangle].
    \label{cpl approx chi2}
\end{equation}
which is precisely the $\Delta \chi^2_{\rm est}$ between CPL and $\rm \Lambda CDM$. So, as expected, CPL and CPL$^*$ yield the same 2-dimensional tension with $\rm \Lambda CDM$ for a given dataset to second order approximation.

Additionally, we may recast \cref{chi opt} from \cref{subsection: optimum} by plugging in the definitions of $a_c$ \cref{ac definition} and $a_p$ \cref{ap definition} to get
\begin{equation}
    \Delta\chi^2_{\rm opt}=\frac{-(w_0+1)^2\langle \delta w_a^2\rangle+2(w_0+1)w_a\langle \delta w_0\delta w_a\rangle-w_a^2\langle \delta w_0^2\rangle}{\langle \delta w_0^2\rangle\langle \delta w_a^2\rangle-\langle \delta w_0\delta w_a\rangle^2}.
\end{equation}
Interestingly, this is equivalent to \cref{cpl approx chi2}. Thus, in the second order approximation of $\log\mathcal{L}$, 
\begin{equation}
    \Delta \chi^2 \approx\Delta \chi_{\rm est}^2=\Delta\chi^2_{\rm opt}.
    \label{chi2 equivalence}
\end{equation}
So, in the case of a Gaussian distribution of the likelihood, the maximum 1-dimensional $\Delta\chi^2$ of $w_*$ and $-1$ is given by the $\Delta \chi^2$ of the CPL$_*$ (or CPL) model with $\rm \Lambda CDM$. The 2-dimensional $\chi^2$ in the $w_0,w_a$ plane, serves as an upper bound on the 1-dimensional tension of $w_*$, thus the optimum captures the entire tension in 2 dimensions. The optimum of a dataset can be used to estimate its tension with $\rm\Lambda CDM$ and vice versa. However, significant non-Gaussianities in the likelihood distribution could cause this approximation to be poor.

\begin{table*}
\begin{threeparttable}
\centering
\footnotesize
\setlength{\tabcolsep}{4pt}
\renewcommand{\arraystretch}{1.2}

\begin{tabular}{@{}lcccc@{}}
\hline\hline
Dataset & Deriv.\ $\sigma_{\rm opt}$ & Deriv.\ $\Delta \chi^2_{\rm opt}$ & Deriv.\ $\Delta \chi^2$ & Samp. $\Delta\chi^2$\\
\hline

\multicolumn{5}{c}{\textit{BAO + CMB}} \\
\hline
DESI2 + CMB & $\mathbf{3.06\ (< -1)}$ & -9.36 & -9.38 & -11.75\\
DESI2 + D6 + CMB & $\mathbf{3.03\ (< -1)}$ & -9.18 & -9.21 & -11.81\\
DESI2 + D6 + CMB + SPT & $\mathbf{3.32\ (< -1)}$ & -11.02 & -11.05 & -15.21\\

\hline
\multicolumn{5}{c}{\textit{BAO + SN + CMB}} \\
\hline
DESI2 + D6 + Dovekie + CMB & $\mathbf{3.63\ (> -1)}$ & -13.18 & -13.19 & -13.58\\
DESI2 + D6 + Union3.1 + CMB & $\mathbf{3.52\ (> -1)}$ & -12.39 & -12.40 & -13.67 \\
DESI2 + D6 + PP\_Hoyt + CMB & $\mathbf{3.62\ (> -1)}$ & -13.10 & -13.07 & -13.19 \\

DESI2 + DESY5 + CMB & 4.45 ($> -1$) & -19.80 & -19.81 & -19.60 \\
DESI2 + Union3 + CMB & 3.97 ($> -1$) & -15.76 & -15.79 & -16.90 \\
DESI2 + PP + CMB & 3.23 ($> -1$) & -10.43 & -10.41 & -10.33 \\

\hline
\multicolumn{5}{c}{\textit{BAO + SN}} \\
\hline
DESI2 + D6 + Dovekie & 2.73 ($> -1$) & -7.45 & -7.43 & -7.36 \\
DESI2 + D6 + Union3.1 & 2.61 ($> -1$) & -6.81 & -6.82 & -6.45 \\
DESI2 + D6 + PP\_Hoyt & 2.60 ($> -1$) & -6.76 & -6.75 & -7.41 \\

DESI2 + DESY5 & 3.86 ($> -1$) & -14.90 & -14.88 & -13.60 \\
DESI2 + Union3 & 3.10 ($> -1$) & -9.61 & -9.63 & -10.08\\
DESI2 + PP & 2.20 ($> -1$) & -4.84 & -4.86 & -4.92 \\

\hline\hline
\end{tabular}

\caption{Derived chi-squared and tension metrics.
Note that derived $\Delta\chi^2$ estimates for both the maximal 1-dimensional tension ($\Delta\chi^2_{\rm opt}$) and the 2-dimensional tension are equivalent within rounding error demonstrating the validity of \cref{chi2 equivalence}. However, some sampled $\Delta\chi^2$ differ notably from derived values (particularly in BAO + CMB dataset combinations).}
\label{tab:derived significance}

\begin{tablenotes}
\item[a] For $a_{\rm opt}>1$, the sampled value corresponds to $w_0$ instead of $w_{\rm opt}$.
\end{tablenotes}

\end{threeparttable}
\end{table*}

\bibliography{Refrences, refs_key_paper, INSPIRE-CiteAll_filtered_sorted, Leo, DESI2024, DDEComment}

\end{document}